\documentclass[onecolumn,fp,seceq]{jpsj3}

\usepackage{txfonts}
\usepackage{graphicx}
\usepackage{dcolumn}
\usepackage{bm}
\usepackage[usenames]{color} 
\usepackage{amssymb} 
\usepackage{amsmath} 
\usepackage[utf8]{inputenc} 
\usepackage{braket}
\usepackage{mathptmx}
\newcommand{\Uell}{u_{\ell {\bm k}}}
\newcommand{\Uellp}{u_{\ell' {\bm k}}}
\newcommand{\Eell}{\varepsilon_{\ell}}
\newcommand{\Eellp}{\varepsilon_{\ell '}}

\newcommand{\Eellx}{\frac{\partial \Eell}{\partial k_x}}
\newcommand{\Eelly}{\frac{\partial \Eell}{\partial k_y}}

\newcommand{\Uelly}{\frac{\partial \Uell}{\partial k_y}}

\title{Orbital Magnetism of Bloch Electrons II. 
Application to Single-Band Models 
and Corrections to Landau-Peierls susceptibility
} 

\author{
Masao \surname{Ogata}
}

\inst{\ 
Department of Physics, University of Tokyo, Hongo, Bunkyo-ku, Tokyo 113-0033, Japan 
} 

\abst{
Orbital susceptibility for Bloch electrons is calculated for the first time  
up to the first order with respect to overlap integrals between the neighboring atomic orbitals, 
assuming single-band models. 
A general and rigorous theory of orbital susceptibility developed in the preceding paper 
is applied to single-band models 
in two-dimensional square and triangular lattices. 
In addition to the Landau-Peierls orbital susceptibility, 
it is found that there are comparable  
contributions from the Fermi surface and from the 
occupied states in the partially filled band called {\it intraband atomic diamagnetism}. 
This result means that the Peierls phase used in tight-binding models is insufficient 
as the effect of magnetic field. 
}

\begin{document}
\maketitle

\section{Introduction}

The effect of magnetic field on electrons in crystals is one of the 
fundamental problems in solid state physics.\cite{Kubo} 
In particular, orbital magnetism and its interband contributions have 
a long history of research.\cite{Peierls,Wilson,Adams,Kjeldaas,Roth,Blount,Wannier,%
Ichimaru,HFKubo,HFKubo2,HS2,HS3,Fukuyama}
However, most preceding calculations have been based on the Landau-Peierls theory, 
which was developed for the single-band tight-binding model.\cite{Peierls}
Calculations of orbital susceptibility based on exact formulae\cite{HS2,HS3} 
for Bloch electrons have not been developed. 

Recently, we have derived an exact formula of orbital susceptibility expressed in terms of 
Bloch wave functions,\cite{OgaFuku1} 
which is simpler than those obtained before\cite{HS2,HS3} and will be useful for explicit calculations. 
We started from the exact one-line formula (Fukuyama formula)\cite{Fukuyama} 
\begin{equation}
\chi = \frac{e^2}{\hbar^2c^2} k_{\rm B} T \sum_{{\bm k},n} {\rm Tr} \ 
\gamma_x {\cal G} \gamma_y {\cal G} \gamma_x {\cal G} \gamma_y {\cal G},
\label{FukuyamaF}
\end{equation}
where $\cal G$ represents the thermal Green's function ${\cal G}({\bm k}, \varepsilon_n)$ 
in a matrix form of band indices,  $\varepsilon_n$ is Matsubara frequency, 
and $\gamma_\mu$ is the current operator in the 
$\mu$-direction divided by $e/\hbar$. 
The spin multiplicity of 2 has been taken into account and Tr means 
to take trace over band indices. 
In our preceding paper\cite{OgaFuku1} (referred to as I in the following),
we rewrote the Fukuyama formula (\ref{FukuyamaF}) in terms of Bloch wave functions 
and obtained a new formula for the orbital susceptibility $\chi$ as follows: 
\begin{equation}
\chi = \chi_{\rm LP} + \chi_{\rm inter} + \chi_{\rm FS} + \chi_{\rm occ},
\label{FinalChi}
\end{equation}
with 
\begin{equation}
\chi_{\rm LP} = \frac{e^2}{6 \hbar^2 c^2} 
\sum_{\ell, {\bm k}} f'(\Eell) 
\left\{ \frac{\partial^2 \Eell}{\partial k_x^2} \frac{\partial^2 \Eell}{\partial k_y^2} 
-\left( \frac{\partial^2 \Eell}{\partial k_x \partial k_y} \right)^2 \right\},
\label{ChiLP}
\end{equation}
\begin{equation}
\begin{split}
\chi_{\rm inter} &= -\frac{e^2}{\hbar^2 c^2} \sum_{\ell \ne \ell', {\bm k}} \frac{f(\Eell)}{\Eell - \Eellp} 
\biggl| \int \frac{\partial \Uell^\dagger}{\partial k_x} 
\left( \frac{\partial H_{\bm k}}{\partial k_y} + \frac{\partial \Eell}{\partial k_y} \right) \Uellp d{\bm r} \cr 
&\qquad\qquad - \int \frac{\partial \Uell^\dagger}{\partial k_y} 
\left( \frac{\partial H_{\bm k}}{\partial k_x} + \frac{\partial \Eell}{\partial k_x} \right) \Uellp 
d{\bm r} \biggr|^2, 
\label{ChiInter}
\end{split}
\end{equation}
\begin{equation}
\begin{split}
\chi_{\rm FS} &= \frac{e^2}{\hbar^2 c^2} \sum_{\ell, {\bm k}} f'(\Eell) \biggl\{ 
\Eellx\int \frac{\partial \Uell^\dagger}{\partial k_y} 
\left( \frac{\partial H_{\bm k}}{\partial k_x} + \frac{\partial \Eell}{\partial k_x} \right)
\Uelly d{\bm r} \cr
&\qquad
-\Eellx\int \frac{\partial \Uell^\dagger}{\partial k_x} 
\left( \frac{\partial H_{\bm k}}{\partial k_y} + \frac{\partial \Eell}{\partial k_y} \right)
\Uelly d{\bm r}\biggr\} + (x\leftrightarrow y),
\label{ChiFS}
\end{split}
\end{equation}
\begin{equation}
\begin{split}
\chi_{\rm occ} &= -\frac{e^2}{2\hbar^2 c^2}
\sum_{\ell, {\bm k}} f(\Eell) \biggl\{ 
\frac{\partial^2 \Eell}{\partial k_x \partial k_y} 
\int \frac{\partial \Uell^\dagger}{\partial k_x} \frac{\partial \Uell}{\partial k_y} d{\bm r} \cr
&\qquad +\left( \frac{\hbar^2}{m} - \frac{\partial^2 \Eell}{\partial k_x^2} \right)
\int \frac{\partial \Uell^\dagger}{\partial k_y} \frac{\partial \Uell}{\partial k_y} d{\bm r}
\biggr\} + (x\leftrightarrow y),
\label{ChiC}
\end{split}
\end{equation}
where $f(\varepsilon)$ is the Fermi distribution function,
$\varepsilon_\ell \equiv \varepsilon_\ell ({\bm k})$ is the $\ell$-th Bloch band energy, and $(x\leftrightarrow y)$ 
represents terms in which $x$ and $y$ are exchanged. 
The suffixes of $\chi_{\rm LP}, \chi_{\rm inter}, \chi_{\rm FS}$, and $\chi_{\rm occ}$ denote  
Landau-Peierls, interband, Fermi surface, and occupied states, respectively.\cite{OgaFuku1}
Here, the range of the real-space integral $\int \cdots d{\bm r}$ has been extended 
to the whole system size by using the periodicity of $\Uell({\bm r})$.\cite{OgaFuku1}
Under the periodic potential $V({\bm r})$, 
wave functions are given by $e^{i{\bm k}\cdot {\bm r}} \Uell({\bm r})$, where 
$\Uell({\bm r})$ satisfies
\begin{equation}
H_{\bm k} \Uell({\bm r}) = \Eell({\bm k}) \Uell({\bm r}),
\label{UellEq}
\end{equation}
with 
\begin{equation}
H_{\bm k} = \frac{\hbar^2 k^2}{2m} - \frac{i\hbar^2}{m} {\bm k}\cdot {\bm \nabla}
- \frac{\hbar^2}{2m} {\bm \nabla}^2 + V({\bm r}).
\label{HamiltonianK}
\end{equation}
Note that the formula in eqs.~(\ref{FinalChi})-(\ref{ChiC}) is exact as eq.~(\ref{FukuyamaF}). 
There are several differences between the formula (\ref{FinalChi})-(\ref{ChiC}) and 
those obtained by Hebborn et al.\cite{HS2,HS3} 
although they are equivalent. 
The detailed comparison is given in I.\cite{OgaFuku1}

It was also found that, in the atomic limit, $\chi_{\rm inter}$ is equal to Van Vleck 
susceptibility and $\chi_{\rm occ}$ is equal to atomic diamagnetism from core-level electrons.\cite{OgaFuku1} 
Then, the band effects on the orbital susceptibility can be calculated 
systematically by studying the effects of overlap integrals between neighboring atomic orbitals 
as a perturbation from the atomic limit. 
Furthermore, it was shown that $\chi_{\rm occ}$ contains contributions not only from the
core-level electrons (i.e., atomic diamagnetism), 
but also from the occupied states in the partially filled band, which we call 
{\it intraband atomic diamagnetism} in this paper. 
This contribution has not been recognized before. 

In this paper, we calculate the orbital susceptibility $\chi$ using eqs.~(\ref{FinalChi})-(\ref{ChiC})
perturbatively with respect to overlap integrals between neighboring atomic orbitals. 
Furthermore, we study single-band models in which only one band crosses the Fermi energy 
and the corresponding band consists of an atomic orbital, i.e., 
the matrix elements with the other orbitals are neglected. 
As examples, the 1s atomic orbital on two-dimensional square and triangular lattices is studied. 
We find that there are several contributions even in this simple model, 
which are not included in previous studies. 
The merit of the present method is that all the contributions to $\chi$ are included. 

The relationship between the tight-binding model and the 
systematic expansion with respect to overlap integrals is worth noting here. 
The hopping integral used in the tight-binding model [and $\Eell({\bm k})$] 
is proportional to the overlap integral. 
As a result, $\chi_{\rm LP}$ in eq.~(\ref{ChiLP}) is in the first order with respect to overlap integrals. 
In this paper, we calculate 
$\chi_{\rm inter}, \chi_{\rm FS}$, and $\chi_{\rm occ}$ 
in eqs.~(\ref{ChiInter})-(\ref{ChiC}) exactly up to the same order with $\chi_{\rm LP}$. 

As shown by Peierls,\cite{Peierls} the effect of the magnetic field can be taken into 
account in tight-binding models by attaching the so-called Peierls phase 
to the hopping integrals. 
Using this Peierls phase, $\chi_{\rm LP}$ is obtained 
{\it in the single-band tight-binding model}.\cite{Peierls} 
This is actually confirmed numerically by Raoux {\it et al.},\cite{Piechon} 
who studied square and triangular lattices. 
However, the above formula (\ref{ChiLP})-(\ref{ChiC}) indicates that 
there are other contributions relating to the deformation of the wave 
functions, i.e., $\partial \Uell/\partial {\bm k}$. 
One may expect that $\chi_{\rm LP}$ is dominant in the single-band model. 
However, as shown in the present paper, 
the other contributions are comparable to $\chi_{\rm LP}$. 
This result means that the Peierls phase used in tight-binding models is insufficient 
as the effect of magnetic field.

This paper is organized as follows. 
In section 2, we develop a formalism of the systematic expansion with respect to the 
overlap integrals starting from the atomic limit using the linear combination of atomic orbitals. 
Then, we calculate the orbital susceptibility in the single-band models 
using 1s atomic orbitals in section 3. 
As examples, we study the square and triangular lattices.  
Section 4 is devoted to summary and discussions. 
Detailed calculations are shown in Appendices.

\section{Orbital susceptibility for single-band models}

In this paper, we calculate the orbital susceptibility $\chi$
using the exact formula (\ref{ChiLP})-(\ref{ChiC}) for single-band models. 
First, we develop the formalism for the first-order perturbation with respect to 
overlap integrals between atomic orbitals. 

\subsection{General formalism of linear combinations of atomic orbitals}

As in I,\cite{OgaFuku1} let us consider a situation in which 
the periodic potential $V({\bm r})$ is written as 
\begin{equation}
V({\bm r}) = \sum_{{\bm R}_i} V_0({\bm r}-{\bm R}_i),
\label{PotSum}
\end{equation}
where ${\bm R}_i$ represents lattice sites and $V_0({\bm r})$ is a potential 
of a single atom. 
In order to construct Bloch wave functions, 
we use the atomic orbitals $\phi_n({\bm r})$ that satisfy 
\begin{equation}
\left( -\frac{\hbar^2}{2m} \nabla^2 + V_0({\bm r}) \right) \phi_n ({\bm r}) = 
E_n \phi_n ({\bm r}). 
\label{phiEq}
\end{equation}
Generally, there is an overlap between neighboring atomic orbitals and 
it is necessary to make orthogonal wave functions.  
In the lowest order with respect to overlap integrals, we obtain\cite{Lowdwin}
\begin{equation}
\Phi_n ({\bm r}-{\bm R}_i) = \phi_n ({\bm r}-{\bm R}_i) - \sum_{j, m} \frac{1}{2} 
s_{ij, nm}^* \phi_{m} ({\bm r}-{\bm R}_j),
\label{OrthoNormal}
\end{equation}
which are orthogonal to each other, and the overlap integral $s_{ij, nm}$ is given by
\begin{equation}
s_{ij, nm} = \int \phi_{n}^*({\bm r}-{\bm R}_i) \phi_{m}({\bm r}-{\bm R}_j) d{\bm r}
-\delta_{i,j} \delta_{n,m}.
\end{equation}
In the following, we calculate orbital susceptibility up to the first order with respect 
to \lq\lq overlap integrals" whose integrand contains the overlap of atomic orbitals, 
$\phi_n^* ({\bm r}-{\bm R}) \phi_m ({\bm r})$ (${\bm R}\ne 0$). 

Using these orthogonal wave functions, 
we consider the linear combination of atomic orbitals (LCAO) 
\begin{equation}
\varphi_{n{\bm k}}^{\rm ortho} ({\bm r}) = \frac{1}{\sqrt{N}}
\sum_{{\bm R}_i} e^{-i{\bm k}({\bm r}-{\bm R}_i)} \Phi_n ({\bm r}-{\bm R}_i),
\label{LCAO}
\end{equation}
as a basis set for $\Uell({\bm r})$. 
Here $N$ is the total number of unit cells.
It is easily shown that $\varphi_{n{\bm k}}^{\rm ortho}({\bm r})$ 
are periodic functions with the same period as $V({\bm r})$. 
Using $\varphi_{n{\bm k}}^{\rm ortho}({\bm r})$, we expand $\Uell({\bm r})$ as  
\begin{equation}
\Uell ({\bm r}) = \sum_n c_{\ell, n}({\bm k}) \varphi_{n{\bm k}}^{\rm ortho} ({\bm r}). 
\end{equation}
The coefficients $c_{\ell, n}({\bm k})$ should be determined in order for $\Uell$ to 
satisfy eq.~(\ref{UellEq}).
This can be achieved by diagonalizing a Hamiltonian whose matrix elements are 
\begin{equation}
h_{nm}({\bm k}) = \int \varphi_{n{\bm k}}^{{\rm ortho}*}({\bm r}) H_{\bm k} 
\varphi_{m{\bm k}}^{\rm ortho} ({\bm r}) d{\bm r}.
\end{equation}
[This formulation is slightly different from that mentioned in I. 
However they are equivalent.]

\subsection{Orbital susceptibility for a single band}

We consider a partially filled single-band model. 
In this model, only the matrix elements between the same atomic orbitals are taken into account.  
Then, the wave function is just
\begin{equation}
\Uell ({\bm r}) = \varphi_{\ell{\bm k}}^{\rm ortho} ({\bm r}),
\label{SingleTBA}
\end{equation} 
and the energy eigenvalue $\varepsilon_\ell({\bm k})$ is given by 
$h_{\ell\ell}({\bm k})$, which can be calculated as 
\begin{equation}
\begin{split}
\varepsilon_\ell({\bm k}) 
&= \frac{1}{N} \sum_{{\bm R}_i, {\bm R}_j} \int e^{i{\bm k}({\bm r}-{\bm R}_j)} 
\Phi_\ell^* ({\bm r}-{\bm R}_j) e^{-i{\bm k}({\bm r}-{\bm R}_i)} \cr
&\quad \times \left\{ -\frac{\hbar^2}{2m} \nabla^2 + V({\bm r}) \right\} 
\Phi_{\ell} ({\bm r}-{\bm R}_i) d{\bm r} \cr
&= \frac{1}{N} \sum_{{\bm R}_i, {\bm R}_j} \int e^{i{\bm k}({\bm R}_i-{\bm R}_j)} 
\Phi_\ell^* ({\bm r}-{\bm R}_j) \cr
&\quad \times \biggl[ \biggl\{ E_\ell + \sum_{{\bm R}_{j'}\ne {\bm R}_i} V_0({\bm r}-{\bm R}_{j'}) 
\biggr\} \phi_{\ell} ({\bm r}-{\bm R}_i) d{\bm r} \cr
&\quad -\sum_{j'} \frac{s_{ij',\ell\ell}^*}{2}  
\biggl\{ E_\ell + \sum_{{\bm R}_{j''}\ne {\bm R}_{j'}} V_0({\bm r}-{\bm R}_{j''}) 
\biggr\} \phi_{\ell} ({\bm r}-{\bm R}_{j'}) d{\bm r} \biggr] \cr
&\equiv E_\ell + \frac{1}{N} \sum_{{\bm R}_i, {\bm R}_j}  e^{i{\bm k}({\bm R}_i-{\bm R}_j)} 
\biggl\{ C_{\ell\ell}({\bm R}_j, {\bm R}_i ) \cr
&\quad -\sum_{j'} \frac{s_{ij',\ell\ell}^*}{2} C_{\ell\ell}({\bm R}_j, {\bm R}_{j'})
-\sum_{j'} \frac{s_{jj',\ell\ell}}{2} C_{\ell\ell}({\bm R}_{j'}, {\bm R}_{i}) \biggr\}, 
\end{split}
\end{equation}
where
\begin{equation}
C_{\ell\ell}({\bm R}_j, {\bm R}_i ) = \int 
\phi_\ell^* ({\bm r}-{\bm R}_j) \sum_{{\bm R}_{j'}\ne {\bm R}_i} 
V_0({\bm r}-{\bm R}_{j'}) \phi_{\ell} ({\bm r}-{\bm R}_i) d{\bm r}. 
\end{equation}
Here we have used the relations in (\ref{PotSum}) and (\ref{phiEq}). 

Up to the first order of overlap integrals, we obtain
\begin{equation}
\begin{split}
\varepsilon_\ell({\bm k})  &= E_\ell + C_{\ell\ell}({\bm R}_i, {\bm R}_i ) 
- \sum_{{\bm R}\ne 0}e^{-i{\bm k}\cdot {\bm R}} t_{\ell\ell}({\bm R}), 
\label{EsingleTBA}
\end{split}
\end{equation}
where ${\bm R}={\bm R}_j-{\bm R}_i$, and 
$t_{\ell\ell}({\bm R})$ represents the hopping integrals used in the tight-binding 
models, which are defined as
\begin{equation}
t_{\ell\ell}({\bm R}) = - C_{\ell\ell}({\bm R}_j, {\bm R}_i )+ \frac{s_{ji,\ell\ell}}{2} \left\{ 
C_{\ell\ell}({\bm R}_j, {\bm R}_j) + C_{\ell\ell}({\bm R}_i, {\bm R}_i) \right\}. 
\label{DefTnm}
\end{equation}
When $V_0({\bm R})$ is long-range, it is difficult to calculate $t_{\ell\ell}({\bm R})$ accurately. 
Here, we assume that 
\begin{equation}
\begin{split}
&\int \phi_\ell^* ({\bm r}-{\bm R}_j) V_0({\bm r}-{\bm R}_{j'}) 
\phi_{\ell} ({\bm r}-{\bm R}_i) d{\bm r} \cr
&\sim \frac{s_{ji,\ell\ell}}{2} \biggl\{
\int \phi_\ell^* ({\bm r}-{\bm R}_j) V_0({\bm r}-{\bm R}_{j'}) 
\phi_{\ell} ({\bm r}-{\bm R}_j) d{\bm r} \cr
&\qquad + 
\int \phi_\ell^* ({\bm r}-{\bm R}_i) V_0({\bm r}-{\bm R}_{j'}) 
\phi_{\ell} ({\bm r}-{\bm R}_i) d{\bm r} \biggr\}, 
\end{split}
\end{equation}
when ${\bm R}_i$ and ${\bm R}_j$ are close to each other, 
and ${\bm R}_{j'}\ne {\bm R}_i, {\bm R}_j$. 
This relation will hold when ${\bm R}_{j'}$ is far away from ${\bm R}_i, {\bm R}_j$, 
and we expect that the difference will be small even if ${\bm R}_{j'}$ is 
close to ${\bm R}_i, {\bm R}_j$. 
Then, the ${\bm R}_{j'}$-summation in eq.~(\ref{DefTnm}) can be evaluated using 
the terms with ${\bm R}_{j'} = {\bm R}_i$ or ${\bm R}_j$, and 
$t_{\ell\ell}({\bm R})$ becomes
\begin{equation}
\begin{split}
&t_{\ell\ell}({\bm R})=
- \int \phi_\ell^* ({\bm r}-{\bm R}_j) V_0({\bm r}-{\bm R}_{j}) 
\phi_{\ell} ({\bm r}-{\bm R}_i) d{\bm r} \cr
&\quad + \frac{s_{ji,\ell\ell}}{2} \biggl\{
\int \phi_\ell^* ({\bm r}-{\bm R}_j) V_0({\bm r}-{\bm R}_{i}) 
\phi_{\ell} ({\bm r}-{\bm R}_j) d{\bm r} \cr
&\qquad\quad + 
\int \phi_\ell^* ({\bm r}-{\bm R}_i) V_0({\bm r}-{\bm R}_{j}) 
\phi_{\ell} ({\bm r}-{\bm R}_i) d{\bm r} \biggr\}.
\label{teffFinal} 
\end{split}
\end{equation}

By substituting eqs.~(\ref{SingleTBA}) and (\ref{EsingleTBA}) into 
eqs.~(\ref{ChiLP})-(\ref{ChiC}), we obtain the orbital susceptibility 
for the single-band model. 
First, $\chi_{\rm LP}$ is the Landau-Peierls susceptibility\cite{Peierls} in which $\Eell({\bm k})$ 
in (\ref{EsingleTBA}) is used. 
Note that the ${\bm k}$-derivatives of $\Eell({\bm k})$ are in the first order of 
overlap integrals, and thus $\chi_{\rm LP}$ is also in the first order. 
For evaluating the other contributions, we use
\begin{equation}
\begin{split}
\frac{\partial \Uell}{\partial k_x} = 
\frac{-i}{\sqrt{N}} \sum_{{\bm R}_i}\left( x-R_{ix} \right)
e^{-i{\bm k}({\bm r}-{\bm R}_i)} \Phi_\ell ({\bm r}-{\bm R}_i),
\end{split}
\end{equation}
with ${\bm R}_i=(R_{ix}, R_{iy}, R_{iz})$.  
Up to the first order of overlap integrals, we obtain
[In the following, we do not show the ${\bm k}$-dependences of $\Eell({\bm k})$ explicitly.]
\begin{equation}
\begin{split}
&\chi_{\rm inter} = -\frac{e^2}{\hbar^2 c^2} \sum_{\ell \ne \ell', {\bm k}} \frac{f(\Eell)}{\Eell - \Eellp} 
\biggl[ \frac{\hbar^2}{m^2} \langle L_z \rangle_{\ell\ell'}  \langle L_z \rangle_{\ell'\ell} 
+ \frac{\hbar}{m} \langle L_z \rangle_{\ell\ell'} \cr 
&\times \biggl\{ \frac{\partial \Eell}{\partial k_y} \langle x \rangle_{\ell'\ell}
-\frac{\partial \Eell}{\partial k_x} \langle y \rangle_{\ell'\ell} 
+\frac{\hbar}{m} \sum_{{\bm R}\ne 0} {\rm e}^{-i{\bm k}{\bm R}} \langle L_z \rangle_{R\ell'\ell} 
\biggr\} + {\rm c.c.} \biggr], 
\label{TBMchiInterX}
\end{split}
\end{equation}
\begin{equation}
\begin{split}
&\chi_{\rm FS} = \frac{e^2}{\hbar^2 c^2} \sum_{\ell, {\bm k}} f'(\Eell) \frac{\partial \Eell}{\partial k_x}
\biggl\{ - \frac{\hbar}{m} \langle L_z y \rangle_{\ell\ell} 
+ \frac{\partial \Eell}{\partial k_x} \langle y^2 \rangle_{\ell\ell} 
- \frac{\partial \Eell}{\partial k_y} \langle xy \rangle_{\ell\ell} \cr
& -\frac{\hbar}{m} \sum_{{\bm R}\ne 0} {\rm e}^{-i{\bm k}{\bm R}} 
\langle L_z y - R_x p_y y + R_y p_x  y \rangle_{R\ell\ell} \biggr\}
+ (x\leftrightarrow y), 
\label{TBMchiFSX}
\end{split}
\end{equation}
and
\begin{equation}
\begin{split}
&\chi_{\rm occ} = -\frac{e^2}{2\hbar^2 c^2} \sum_{\ell, {\bm k}} f(\Eell)
\biggl\{ \left( \frac{\hbar^2}{m} 
-\frac{\partial^2 \Eell}{\partial k_x^2} \right) \langle y^2 \rangle_{\ell\ell} \cr
&+ \frac{\partial^2 \Eell}{\partial k_x \partial k_y} 
\langle xy \rangle_{\ell\ell}
+\frac{\hbar^2}{m} \sum_{{\bm R}\ne 0} {\rm e}^{-i{\bm k}{\bm R}} 
\langle (y-R_y)y \rangle_{R\ell\ell} \biggr\}  + (x\leftrightarrow y) , 
\label{ChiExpand2}
\end{split}
\end{equation}
where $\bm R$ is defined as ${\bm R}={\bm R}_j-{\bm R}_i=(R_x, R_y, R_z)$ and 
the expectation values for an operator $\cal O$ are given by
\begin{equation}
\begin{split}
\langle {\cal O} \rangle_{\ell\ell'} 
&= \int \Phi_\ell^* ({\bm r}) {\cal O} \Phi_{\ell'} ({\bm r}) d{\bm r}, \cr
\langle {\cal O} \rangle_{R\ell\ell'} 
&= \int \Phi_\ell^* ({\bm r}-{\bm R}) {\cal O} \Phi_{\ell'} ({\bm r}) d{\bm r}.
\label{Odefinition}
\end{split}
\end{equation}
$\bm R$-summations in eqs.~(\ref{TBMchiInterX})-(\ref{ChiExpand2}) 
come from the integrals between the different sites, which are in the 
first order of overlap integrals.

\section{Application to the 1s orbital case}

To calculate the orbital susceptibility $\chi$ explicitly, 
we assume a simple Coulomb potential for $V_0({\bm r})$, i.e., 
$V_0({\bm r})=-e^2/r$ and 1s orbital for $\phi_\ell({\bm r})$  
\begin{equation}
\phi_{\rm 1s} ({\bm r}) = \frac{1}{\sqrt{\pi} a_{\rm B}^{3/2}} e^{-r/a_{\rm B}}.
\label{1sOrbital}
\end{equation}
Here $a_{\rm B}$ is the Bohr radius $a_{\rm B}=\hbar^2/me^2$. 
We assume that the 1s-orbital band is partially filled and 
only $\ell=$1s is considered. 

First, let us consider $t_{\ell\ell}({\bm R})=t_{\rm 1s1s}({\bm R})$ and 
$\varepsilon_{\rm 1s}({\bm k})$ calculated from (\ref{EsingleTBA}) and (\ref{teffFinal}). 
Since $V_0({\bm r})$ and 1s orbital are isotropic, integrals in 
$t_{\rm 1s1s}({\bm R})$ are independent of the direction of $\bm R$. 
Considering nearest-neighbor sites, we obtain
\begin{equation}
\varepsilon_{\rm 1s}({\bm k}) = E_{\rm 1s} + C_{\rm 1s1s} + \epsilon_{\bm k}, 
\label{SingleTBAene}
\end{equation}
where 
\begin{equation}
\epsilon_{\bm k} = -t \gamma_{\bm k}, 
\end{equation}
and 
\begin{equation}
\gamma_{\bm k} = \sum_{{\bm R}={\rm n.n.}} e^{-i{\bm k}\cdot {\bm R}},
\label{DefGamma}
\end{equation}
with
\begin{equation}
\begin{split}
t &= t_0 + s c_{\rm 1s}, \cr
t_0 &=  -\int \phi_{\rm 1s}^* ({\bm r}-{\bm R}) V_0({\bm r}-{\bm R}) \phi_{\rm 1s} ({\bm r}) d{\bm r}, \cr
s &= \int \phi_{\rm 1s}^*({\bm r}-{\bm R}) \phi_{\rm 1s}({\bm r}) d{\bm r}, \cr
c_{\rm 1s} &=\int \phi_{\rm 1s}^* ({\bm r}) V_0({\bm r}-{\bm R}) \phi_{\rm 1s} ({\bm r}) d{\bm r}. 
\label{tdef1s}
\end{split}
\end{equation}
The ${\bm R}$-summation in $\gamma_{\bm k}$ represents the summation 
over the nearest-neighbor (n.n.) sites. 
(Here, we have assumed only the nearest-neighbor hopping integrals, 
but the extension to the longer-range hopping integrals is straightforward.) 
In the following, the constant energy $E_{\rm 1s} + C_{\rm 1s1s}$ is 
included in the chemical potential $\mu$ and we write 
$f(\varepsilon_{\bm k})$ in place of $f(E_{\rm 1s} + C_{\rm 1s1s}+ \varepsilon_{\bm k})$ 
for simplicity. 

For the 1s orbital and $V_0({\bm r})=-e^2/r$, the integrals in $t$ are analytically 
calculated as\cite{Mulliken} (see Appendix A)
\begin{equation}
\begin{split}
t_0 &=  \frac{e^2}{a_{\rm B}} \left( 1+p \right)\ e^{-p} 
= \frac{\hbar^2}{m a_{\rm B}^2} \left( 1+p \right)\ e^{-p}, \cr
s &= \left( 1+p+\frac{p^2}{3} \right)\ e^{-p},  \cr
c_{\rm 1s} &= -\frac{e^2}{a_{\rm B}} \left\{ \frac{1}{p} - \left(1+\frac{1}{p} \right) e^{-2p} \right\}
\sim -\frac{\hbar^2}{m a_{\rm B}^2} \frac{1}{p},
\label{OverlapInt1s}
\end{split}
\end{equation}
with $p=a/a_{\rm B}$ and $a$ being the distance between the n.n.\ sites, 
i.e., $a=|{\bm R}|$. 
Figure 1 shows the $p$-dependences of $s, t_0$, and $t$. 
Since $s$ should be a small parameter, we choose $p>4$ in the following. 

\begin{figure}
\hspace{2.5truecm}
\includegraphics[width=9cm]{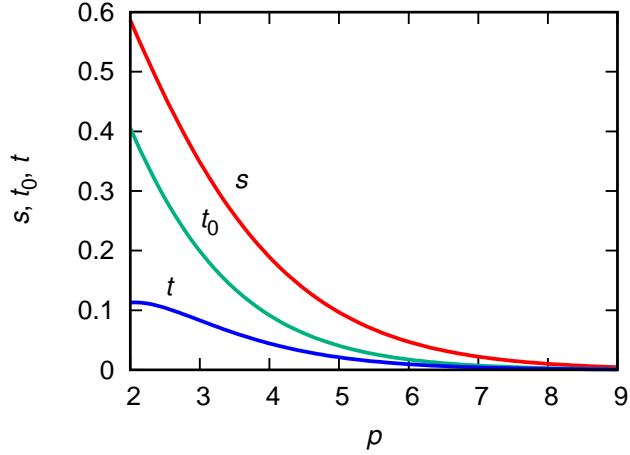}
\vskip 2.5truecm
\caption{(Color online) Overlap integral $s$ and hopping integral $t$ for 1s atomic orbital 
defined in eq.~(\ref{tdef1s}) as a function of the atomic 
distance normalized by the Bohr radius $a_{\rm B}$, i.e., $p=a/a_{\rm B}$.  
For comparison, $t_0$ is also shown.
$t$ and $t_0$ are in the unit of $\hbar^2/ma_{\rm B}^2$. 
}
\label{Fig:Overlap}
\end{figure}

Using these expressions, we obtain 
\begin{equation}
\chi_{\rm LP} = \frac{e^2}{6 \hbar^2 c^2} 
\sum_{{\bm k}} f'(\varepsilon_{\bm k}) 
\left( \varepsilon_{xx} \varepsilon_{yy} - \varepsilon_{xy}^2 \right),
\label{ChiLPSB}
\end{equation}
where we have used abbreviations as 
\begin{equation}
\varepsilon_x = \frac{\partial \varepsilon_{\bm k}}{\partial k_x},  \quad
\varepsilon_{xx} = \frac{\partial^2 \varepsilon_{\bm k}}{\partial k_x^2}, \quad 
\varepsilon_{xy} = \frac{\partial^2 \varepsilon_{\bm k}}{\partial k_x \partial k_y}, \ {\rm etc.}
\label{AbbrevE}
\end{equation}
Expectation values in $\chi_{\rm inter}, \chi_{\rm FS}$, and $\chi_{\rm occ}$
in eqs.~(\ref{TBMchiInterX})-(\ref{ChiExpand2}) should be 
carefully calculated since $\Phi_{\rm 1s} ({\bm r})$ is different from 
$\phi_{\rm 1s} ({\bm r})$ as 
\begin{equation}
\Phi_{\rm 1s} ({\bm r}-{\bm R}_i) = \phi_{\rm 1s} ({\bm r}-{\bm R}_i) - \sum_{j={\rm n.n.}} \frac{s}{2} 
\phi_{\rm 1s} ({\bm r}-{\bm R}_j). 
\label{OrthoNormal1s}
\end{equation}
[See eq.~(\ref{OrthoNormal})]. 
Detailed calculations of the expectation values are shown in Appendix A. 
In particular, owing to the isotropy of the 1s orbital, 
several matrix elements such as $\langle xy\rangle_{\rm 1s1s}$ and 
$\langle xp_y\rangle_{\rm 1s1s}$ vanish. 
Furthermore, we obtain
\begin{equation}
\chi_{\rm inter}=0,
\end{equation}
because of $L_z \phi_{\rm 1s}({\bm r})=0$. 

Next, by using 
(\ref{FormulaRll})
and $\langle x^2   \rangle_{\rm 1s1s} = \langle y^2 \rangle_{\rm 1s1s} = a_{\rm B}^2$,
$\chi_{\rm FS}$ in eq.~(\ref{TBMchiFSX}) becomes
\begin{equation}
\begin{split}
&\chi_{\rm FS} = \frac{e^2}{\hbar^2 c^2} \sum_{\ell, {\bm k}} f'(\Eell) \varepsilon_x
\left( \varepsilon_x a_{\rm B}^2 +\frac{\hbar}{m} \sum_{{\bm R}\ne 0} {\rm e}^{-i{\bm k}{\bm R}} 
\frac{i\hbar}{4} sR_x \right) + (x\leftrightarrow y).
\label{AppendixB0}
\end{split}
\end{equation}
The $\bm R$-summation can be carried out using
\begin{equation}
\sum_{{\bm R}\ne 0} iR_x {\rm e}^{-i{\bm k}{\bm R}} 
= - \frac{\partial}{\partial k_x} \sum_{{\bm R}\ne 0} {\rm e}^{-i{\bm k}{\bm R}}
= - \frac{\partial \gamma_{\bm k}}{\partial k_x} 
= \frac{\varepsilon_{x}}{t},
\end{equation}
where the definition of $\gamma_{\bm k}$ in eq.~(\ref{DefGamma}) is used.
As a result, we obtain
\begin{equation}
\begin{split}
\chi_{\rm FS} 
&= \frac{e^2}{\hbar^2 c^2} \sum_{{\bm k}} f'(\varepsilon_{\bm k}) 
\left( a_{\rm B}^2+\frac{\hbar^2 s}{4m t} \right)  (\varepsilon_{x}^2 + \varepsilon_y^2 ) \cr
&= \frac{e^2}{\hbar^2 c^2} \left( 1+b_1 \right) 
\sum_{{\bm k}} f'(\varepsilon_{\bm k}) a_{\rm B}^2  (\varepsilon_{x}^2 + \varepsilon_y^2 ), 
\label{TBMchiFSSB2}
\end{split}
\end{equation}
where $b_1$ is defined as
\begin{equation}
b_1 = \frac{\hbar^2 s}{4mt a_{\rm B}^2} =  
\frac{1+p+\frac{p^2}{3}}{4\left\{ 1+p -\frac{1}{p} (1+p+\frac{p^3}{3}) \right\}}.
\end{equation}
Similarly, we calculate $\chi_{\rm occ}$ in eq.~(\ref{ChiExpand2}) 
using (\ref{FormulaRll}) and (\ref{x2in1s}) and obtain
\begin{equation}
\begin{split}
\chi_{\rm occ} &= -\frac{e^2}{2\hbar^2 c^2} \sum_{{\bm k}} f(\varepsilon_{\bm k})
\biggl[ \left( \frac{2\hbar^2}{m}- \varepsilon_{xx}- \varepsilon_{yy}\right) a_{\rm B}^2  \cr
&+\frac{\hbar^2}{m} \sum_{{\bm R}\ne 0} {\rm e}^{-i{\bm k}{\bm R}} 
\left( \frac{2a^2}{15} (1+p)\ e^{-p} - \frac{R_x^2 + R_y^2}{5} s \right) \biggr].
\label{AppendixB1}
\end{split}
\end{equation}
Again, the $\bm R$-summation can be carried out using
\begin{equation}
\begin{split}
\sum_{{\bm R}\ne 0} R_x^2 {\rm e}^{-i{\bm k}{\bm R}} 
= - \frac{\partial^2}{\partial k_x^2} \sum_{{\bm R}\ne 0} {\rm e}^{-i{\bm k}{\bm R}}
= - \frac{\partial^2 \gamma_{\bm k}}{\partial k_x^2} 
= \frac{\varepsilon_{xx}}{t},
\end{split}
\end{equation}
Substituting this result into (\ref{AppendixB1}), we obtain 
\begin{equation}
\begin{split}
\chi_{\rm occ} 
&= -\frac{e^2}{2\hbar^2 c^2} \sum_{{\bm k}} f(\varepsilon_{\bm k})
\biggl[ \frac{2\hbar^2}{m} a_{\rm B}^2 - b_2 a^2 a_{\rm B}^2\varepsilon_{\bm k}  \cr
&\qquad\qquad - \left( 1+ \frac{4}{5} b_1 \right) 
a_{\rm B}^2 (\varepsilon_{xx}+\varepsilon_{yy}) \biggr] \cr 
&\equiv \chi_{{\rm occ}:1} + \chi_{{\rm occ}:2} + \chi_{{\rm occ}:3},
\label{TBMchioccSB}
\end{split}
\end{equation}
where the $j$-th term in $\chi_{{\rm occ}}$ is denoted as $\chi_{{\rm occ}:j}$ and 
$b_2$ is defined as
\begin{equation}
b_2 =\frac{2}{15} \frac{\hbar^2 (1+p)\ e^{-p}}{mt a_{\rm B}^2} 
=\frac{2(1+p)}{15 \left\{ 1+p - \frac{1}{p} (1+p+\frac{p^3}{3}) \right\}}.
\end{equation}

The above results are valid in two- and three-dimensions.
There are several remarks. 

(1) The first term in $\chi_{\rm occ}$, i.e., $\chi_{{\rm occ}:1}$, 
does not depend on the overlap integral except for $f(\varepsilon_{\bm k})$. 
As discussed in I\cite{OgaFuku1}, this is a contribution from the occupied 
states in the partially filled band (in this case, the 1s band), which 
we call {\it intraband atomic diamagnetism} in this paper. 
This term is proportional to the electron number in the band, i.e.,
\begin{equation}
\chi_{{\rm occ}:1} = -\frac{e^2 a_{\rm B}^2}{2m c^2} n(\mu),
\end{equation}
where $n(\mu)$ represents the total electron number with the spin degeneracy 
when the chemical potential is $\mu$. 
The other terms $\chi_{{\rm occ}:2}$ and $\chi_{{\rm occ}:3}$ as well as $\chi_{\rm LP}$ and 
$\chi_{\rm FS}$ are in the first order of overlap integrals, i.e., proportional to $e^{-p}$. 

(2) The last term $\chi_{{\rm occ}:3}$ can be rewritten as
\begin{equation}
\begin{split}
&\frac{e^2}{2\hbar^2 c^2} \left( 1+\frac{4}{5} b_1 \right) 
 \sum_{{\bm k}} f(\varepsilon_{\bm k})a_{\rm B}^2 (\varepsilon_{xx}+\varepsilon_{yy})  \cr
&=- \frac{e^2}{2\hbar^2 c^2} \left( 1+\frac{4}{5} b_1 \right) 
 \sum_{{\bm k}} f'(\varepsilon_{\bm k})a_{\rm B}^2 (\varepsilon_{x}^2+\varepsilon_{y}^2),
\label{OccToFS}
\end{split}
\end{equation}
by integration by parts.
We can see that this term is approximately half of $\chi_{\rm FS}$ in (\ref{TBMchiFSSB2}) 
with an opposite sign. 

(3) When the 1s-orbital band is fully filled, 
$\chi_{\rm LP}=\chi_{\rm FS}=0$ owing to the absence of the Fermi surface. 
Furthermore, the $\bm k$-summation in
$\chi_{\rm occ}$ becomes the sum over the whole Brillouin zone. 
In this case, we can see that $\chi_{{\rm occ}:2}$ and $\chi_{{\rm occ}:3}$ vanish. 
As a result, only $\chi_{{\rm occ}:1}$ contributes to 
the orbital susceptibility, i.e.,
\begin{equation}
\chi = -\frac{e^2 a_{\rm B}^2}{mc^2} N,
\label{AtomicDia1s}
\end{equation}
which is nothing but the atomic diamagnetism from the 1s core electrons.
This means that the dispersion $\varepsilon_{\bm k}$ due to the finite overlap 
between the neighboring atomic orbitals  
does not lead to a modification of the atomic diamagnetism. 

In order to calculate the numerical coefficients and compare the magnitude 
of each term, we need to assume a certain lattice structure. 
In the following subsections, we study square lattice and triangular lattice, 
as examples. 

\subsection{Square lattice}

In the case of the two-dimensional square lattice, we have 
\begin{equation}
\varepsilon_{\bm k} =-t\gamma_{\bm k}  = -2t \left( \cos k_x a + \cos k_y a \right).
\end{equation}
Therefore, a simple relation 
$\varepsilon_{xx} + \varepsilon_{yy}=-a^2 \varepsilon_{\bm k}$ 
holds. 
Actually, we find that this relation holds in every two-dimensional lattice 
with nearest-neighbor hopping, because
\begin{equation}
\varepsilon_{xx} + \varepsilon_{yy}
=-t \left( \frac{\partial^2 \gamma_{\bm k}}{\partial k_x^2} + \frac{\partial^2 \gamma_{\bm k}}{\partial k_y^2} \right) 
= t \sum_{\bm R} (R_x^2+R_y^2) e^{-i{\bm k}{\bm R}} 
= -a^2 \varepsilon_{\bm k}. 
\end{equation}
Using this relation, we obtain at $T=0$
\begin{equation}
\chi_{\rm LP} = -\frac{2e^2}{3\hbar^2 c^2} {ta^4} 
\sum_{{\bm k}} \delta\left( (\varepsilon_{\bm k}-\mu)/t \right) \cos k_x a \cos k_y a,
\label{ChiSquareLPint}
\end{equation}
\begin{equation}
\begin{split}
\chi_{\rm FS} &= -\frac{4e^2}{\hbar^2 c^2} t a^2 a_{\rm B}^2  
\left( 1 + b_1 \right) \cr
&\qquad \times \sum_{{\bm k}} \delta\left( (\varepsilon_{\bm k}-\mu)/t \right)
 (\sin^2 k_x a + \sin^2 k_y a ), 
 \label{ChiSquareFSint}
\end{split}
\end{equation}
and
\begin{equation}
\begin{split}
\chi_{\rm occ} &= -\frac{e^2 a_{\rm B}^2}{2m c^2} n(\mu) 
+\frac{2e^2}{\hbar^2 c^2} t a^2 a_{\rm B}^2 
\left( 1-b_2 + \frac{4}{5} b_1\right) \cr
&\qquad \times \sum_{{\bm k}} \delta\left( (\varepsilon_{\bm k}-\mu)/t \right)
 (\sin^2 k_x a + \sin^2 k_y a ), 
\end{split}
\end{equation}
where we have used the integration by parts in (\ref{OccToFS}).

$\bm k$-summations in the thermodynamic limit can be carried out and 
expressed by elliptic integrals as follows (see details in Appendix B):
\begin{equation}
\chi_{\rm LP} = -\frac{4}{3\pi} \left\{ E(k) - \frac{1}{2} K(k) \right\} \chi_0, 
\label{SquareLPanal}
\end{equation}
\begin{equation}
\chi_{\rm FS} = -\frac{16}{\pi} \frac{1}{p^2} 
\left( 1 + b_1 \right) \left\{ E(k) - (1-k^2) K(k) \right\} \chi_0, 
\end{equation}
and
\begin{equation}
\begin{split}
\chi_{\rm occ} &= -\frac{e^2 a_{\rm B}^2}{2m c^2} n(\mu) \cr
&+ \frac{8}{\pi} \frac{1}{p^2} \left( 1-b_2+\frac{4}{5} b_1 \right) 
\left\{ E(k) - (1-k^2) K(k) \right\} \chi_0,
\end{split}
\end{equation}
where $K(k)$ ($E(k)$) is the complete elliptic integral of the first (second) kind with 
$k=\sqrt{1-\mu^2/16t^2}$, and 
$\chi_0$ is the Pauli susceptibility at the bottom of the band ($\mu=-4t$) given by
\begin{equation}
\chi_0 = \frac{e^2}{2\pi \hbar^2 c^2} t a^2 L^2,
\end{equation}
with $L^2$ being the system size ($L^2=a^2N$). 
Here, we have used the fact that the model is equivalent to free electrons with 
an effective mass $m^*=\hbar^2/2t a^2$ at the bottom of the band. 
The electron number $n(\mu)$ can be calculated from 
$n(\mu)/L^2=2\int^\mu_{-4t} D(\mu) d\mu$ with $D(\mu)$ being the density of states 
per area for the two-dimensional square lattice: (see Appendix B)
\begin{equation}
D(\mu) = \frac{1}{2\pi^2 t a^2} K(k).
\end{equation}

\begin{figure}
\hspace{0.2truecm}
\includegraphics[width=13.333cm]{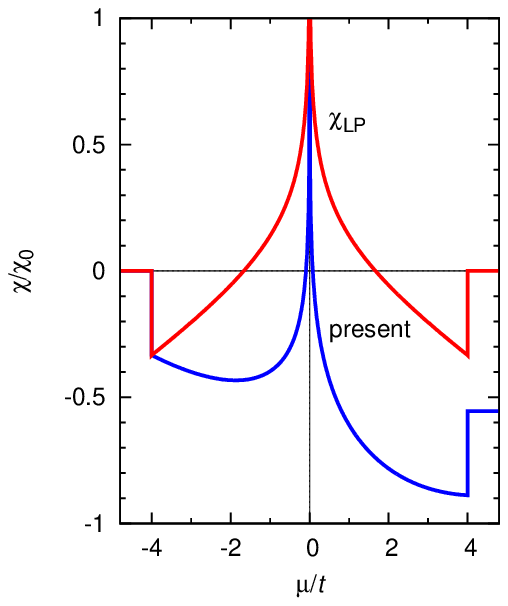}

\hspace{0.6truecm}
\includegraphics[width=12cm]{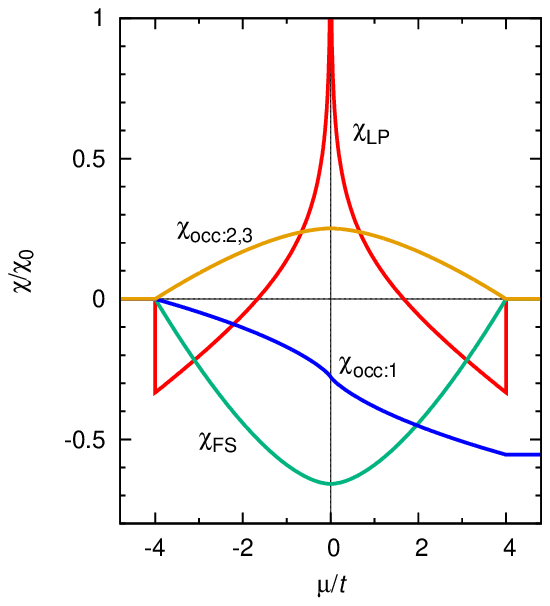}
\vskip 2truecm
\caption{(Color online) (a) Orbital susceptibility as a function of chemical potential $\mu$ in the 
case of two-dimensional square lattice, normalized by the Pauli susceptibility 
$\chi_0$ at the band edge. 
For comparison, Landau-Peierls susceptibility $\chi_{\rm LP}$ is also shown. 
(b) Each contribution, $\chi_{\rm LP}, \chi_{\rm FS}, \chi_{{\rm occ}:1}$, and 
$\chi_{{\rm occ}:2,3}\equiv \chi_{{\rm occ}:2}+\chi_{{\rm occ}:3}$, as a function of $\mu$.}
\label{Fig:SquareChi}
\end{figure}

Figure \ref{Fig:SquareChi}(a) shows the obtained susceptibility as a function of $\mu$, 
compared with $\chi_{\rm LP}$ when $p\equiv a/a_{\rm B}=4$ as a typical case. 
Apparently, there is a sizable difference from $\chi_{\rm LP}$ even in this 
simple single-band model. 
Furthermore, an asymmetry with respect to the sign change of $\mu$ appears 
in the present result. 
In order to understand this total $\chi$, 
each contribution is shown in Fig.~\ref{Fig:SquareChi}(b) as a function of $\mu$. 
Following are several remarks on these results. 

(1) $\chi_{\rm LP}$ is equal to $-1/3\chi_0$ at the band edge ($\mu=\pm 4t$), 
which is understood as the Landau orbital susceptibility for free electrons. 
When $\mu=\pm 4t$, $k$ in the elliptic integrals is equal to 0, and thus $K(0)=E(0)=\pi/2$. 
As a result, it is confirmed that $\chi_{\rm LP}=-1/3\chi_0$ in eq.~(\ref{SquareLPanal}).
As shown in Fig.~\ref{Fig:SquareChi}, $\chi_{\rm LP}$ increases as $\mu$ increases, 
crosses zero at $\mu=-1.667t$, and 
has a diverging peak at $\mu=0$, which is a well-known behavior.\cite{Piechon} 
This divergence corresponds to the van Hove singularities at ${\bm k}=(\pi,0)$ and $(0,\pi)$, 
and it is analytically given by 
\begin{equation}
\chi_{\rm LP}(\mu\rightarrow 0) \sim \frac{2}{3\pi} \ln \left( \frac{16t}{|\mu|} \right) \chi_0,  
\end{equation}
from eq.~(\ref{SquareLPanal}). 
Here, we have used $K(k) \sim \ln(4/\sqrt{1-k^2})$ as $k\rightarrow 1$. 
This divergence is $\frac{2}{3} \frac{e^2}{\hbar^2 c^2} a^4 t^2 L^2$ times larger 
than the divergence of the density of states, $D(\mu)$, which is reasonable 
since the integrand in (\ref{ChiSquareLPint}) is $\cos k_xa \cos k_y a =-1$ 
at the van Hove singularities ${\bm k}=(\pi,0)$ and $(0,\pi)$. 

(2) $\chi_{\rm FS}$ is always negative and has its maximum absolute value at $\mu=0$. 
There is no divergence at the van Hove singularity because the integrand $\sin^2 k_x a + \sin^2 k_y a$ 
in (\ref{ChiSquareFSint}) vanishes at ${\bm k}=(\pi,0)$ and $(0,\pi)$. 
As shown in Fig.~\ref{Fig:SquareChi}(b), $\chi_{\rm FS}$ is comparable to $\chi_{\rm LP}$. 

(3) Among three contributions in $\chi_{\rm occ}$, $\chi_{{\rm occ}:1}$ is the intraband atomic 
diamagnetism, which is asymmetric with respect to $\pm \mu$. 
This causes the asymmetry of the total $\chi$, as shown in Fig.~\ref{Fig:SquareChi}(a). 
When the band is fully occupied (i.e., $\mu>4t$), only $\chi_{{\rm occ}:1}$ remains, which is the same 
as the atomic diamagnetism of the 1s band. 
On the other hand, $\chi_{{\rm occ}:2,3} \equiv \chi_{{\rm occ}:2}+\chi_{{\rm occ}:3}$ 
is positive and approximately cancels with half of $\chi_{\rm FS}$, as discussed before. 

\begin{figure}
\hspace{1.5truecm}
\includegraphics[width=10cm]{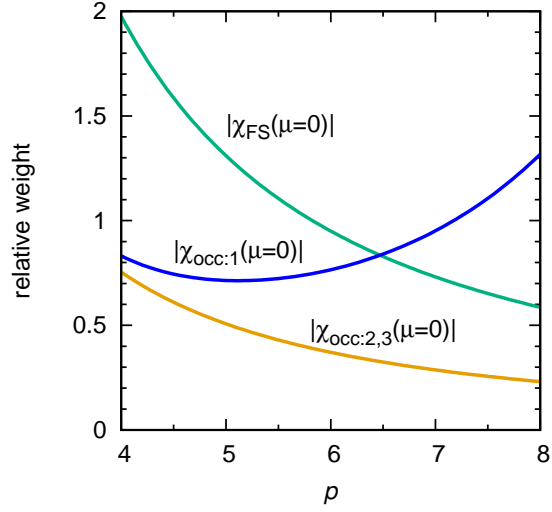}
\vskip 2truecm
\caption{(Color online) Relative weight of each component of $\chi$ at a typical value of $\mu=0$ 
normalized by $|\chi_{\rm LP}(\mu=-4t)|$ as a function of $p=a/a_{\rm B}$.}
\label{Fig:SquareRel}
\end{figure}

Figure \ref{Fig:SquareChi} is the result for a typical case with $p=a/a_{\rm B}=4$. 
In order to see the relative weight of each contribution more closely, we 
study the $p$-dependence of each contribution at a special value of $\mu$. 
For $\chi_{\rm LP}$, we use the value at $\mu=-4t$, which is $-1/3$ of $\chi_0$. 
For $\chi_{\rm FS}, \chi_{{\rm occ}:1}$, and $\chi_{{\rm occ}:2,3}$ we use the value at 
$\mu=0$ as a typical case, i.e., 
\begin{equation}
\begin{split}
\chi_{\rm FS}(\mu=0) &= -\frac{16}{\pi} \frac{1}{p^2} 
\left( 1 + b_1 \right) \chi_0, \cr
\chi_{{\rm occ}:1} (\mu=0) &= -\frac{e^2 a_{\rm B}^2}{2\hbar^2 c^2} N \cr
&= -\frac{\pi e^p}{p^4 \left\{ 1+p - \frac{1}{p} (1+p+\frac{p^3}{3}) \right\}} \chi_0, \cr
\chi_{{\rm occ}:2,3} (\mu=0) &= \frac{8}{\pi} \frac{1}{p^2} 
\left( 1-b_2+\frac{4}{5} b_1 \right) \chi_0,
\label{TypicalOcc2}
\end{split}
\end{equation}
where we have used the expression of $t$ in eq.~(\ref{tdef1s}).  
Figure \ref{Fig:SquareRel} shows the relative weights 
of $|\chi_{\rm FS}|$, $|\chi_{{\rm occ}:1}|$, and $|\chi_{{\rm occ}:2,3}|$ against $|\chi_{\rm LP}|$
as a function of $p=a/a_{\rm B}$. 
We can see that the relative weights for 
$|\chi_{\rm FS}|$ and $|\chi_{{\rm occ}:2,3}|$ become smaller as $p$ increases. 
This is mainly due to their numerical prefactor $1/p^2$ in (\ref{TypicalOcc2}), 
whose origin is that $\chi_{\rm LP}$ has a factor $(a/a_{\rm B})^4=p^4$ 
owing to the 4 times $\bm k$-derivatives in $\varepsilon_{xx} \varepsilon_{yy}-\varepsilon_{xy}^2$, 
whereas  $\chi_{\rm FS}$ and $\chi_{{\rm occ}:2,3}$ have $(a/a_{\rm B})^2=p^2$ from 
$\varepsilon_{x}^2+ \varepsilon_{y}^2$ in $\chi_{\rm FS}$ or 
$\varepsilon_{xx} + \varepsilon_{yy}$ in $\chi_{{\rm occ}:2,3}$. 
As a result, $\chi_{\rm LP}$ becomes dominant as $p$ increases. 

On the other hand, the relative weight of $|\chi_{{\rm occ}:1}|$ increases as $p$ increases. 
The reason for this is as follows: 
Since $\chi_{{\rm occ}:1}$ is the intraband atomic diamagnetism, it does not depend on $p$, 
while $\chi_{\rm LP}\propto t$ decays exponentially as a function of $p$. 
As a result, the contribution of $\chi_{{\rm occ}:1}$ becomes important as $p$ increases, 
which was not recognized before.

\subsection{Triangular lattice}

The application to the two-dimensional triangular lattice is straightforward. 
In this case, we have 
\begin{equation}
\varepsilon_{\bm k} =-t\gamma_{\bm k}  = -2t 
\left( \cos k_x a + 2\cos \frac{k_x a}{2} \cos \frac{\sqrt{3}k_y a}{2} \right), 
\end{equation}
where $a$ is the distance between the nearest-neighbor sites. 
Again, 
$\varepsilon_{xx} + \varepsilon_{yy}=-a^2 \varepsilon_{\bm k}=ta^2 \gamma_{\bm k}$
holds as discussed in the previous subsection. 

Using this dispersion relation, the $\bm k$-summations are carried out analytically 
at $T=0$ and we obtain 
\begin{equation}
\chi_{\rm LP} = -\frac{2\pi}{9} \frac{I_1(\mu)}{a^2t} \chi_0, 
\label{TriLPanal}
\end{equation}
\begin{equation}
\chi_{\rm FS} = -\frac{4\pi}{3} \frac{1}{p^2} 
\left( 1 + b_1 \right) \frac{I_2(\mu)}{t} \chi_0, 
\end{equation}
and
\begin{equation}
\chi_{\rm occ} = -\frac{e^2 a_{\rm B}^2}{2m c^2} n(\mu) + \frac{2\pi}{3} \frac{1}{p^2} 
\left( 1-b_2+\frac{4}{5} b_1 \right) \frac{I_2(\mu)}{t} \chi_0,
\end{equation}
where the analytical forms of $I_1(\mu)$ and $I_2(\mu)$ are shown in Appendix C using 
the elliptic integrals. 
$\chi_0$ represents the Pauli susceptibility 
at the bottom of the band ($\mu=-6t$)
\begin{equation}
\chi_0 = \frac{3e^2}{4\pi \hbar^2 c^2} ta^2 L^2.
\end{equation}
$n(\mu)$ can be calculated as $n(\mu)/L^2=2\int_{-6t}^\mu D(\mu)d\mu$, with 
the density of states per area (see Appendix C)
\begin{equation}
\begin{split}
D(\mu) &= \frac{1}{\sqrt{3} \pi^2 ta^2} \frac{1}{\sqrt{\eta}} K(\kappa), \quad {\rm for}\ -6<\frac{\mu}{t}<2, \cr
D(\mu) &= \frac{1}{\sqrt{3} \pi^2 ta^2} \frac{1}{\kappa\sqrt{\eta}} K(\frac{1}{\kappa}), \quad {\rm for}\ 2<\frac{\mu}{t}<3,
\label{TriangleDOS}
\end{split}
\end{equation}
with $\kappa=\sqrt{(-\mu^2/t^2+12+8\eta)/\eta}$ and $\eta=\sqrt{3-\mu/t}$. 

\begin{figure}
\hspace{1truecm}
\includegraphics[width=12cm]{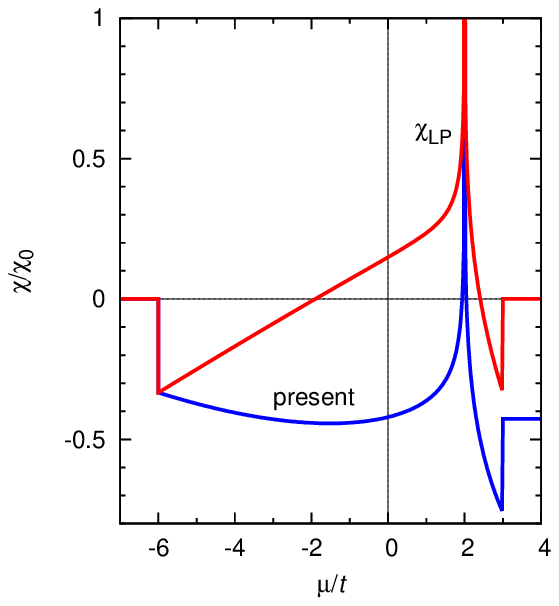}

\hspace{1truecm}
\includegraphics[width=12cm]{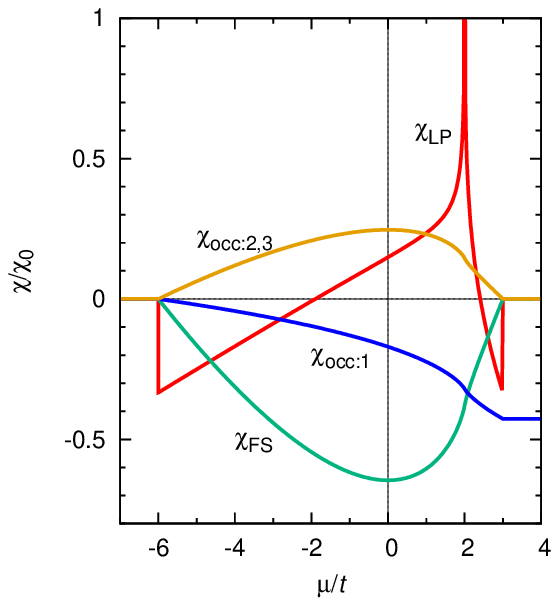}
\vskip 2truecm
\caption{(Color online) (a) Orbital susceptibility as a function of chemical potential $\mu$ in the 
case of two-dimensional triangular lattice, normalized by the Pauli susceptibility 
$\chi_0$ at the band edge. 
For comparison, Landau-Peierls susceptibility $\chi_{\rm LP}$ is also shown. 
(b) Each contribution, $\chi_{\rm LP}, \chi_{\rm FS}, \chi_{{\rm occ}:1}$, and 
$\chi_{{\rm occ}:2,3}$, as a function of $\mu$.}
\label{Fig:TriangleChi}
\end{figure}

In Fig.~\ref{Fig:TriangleChi}, we show the obtained susceptibility as a function of 
$\mu$ for $p=a/a_{\rm B}=4$. 
The behavior is similar to the square-lattice case shown in Fig.~\ref{Fig:SquareChi}, 
i.e., there is a sizable difference from $\chi_{\rm LP}$. 
There are several remarks.

(1) $\chi_{\rm LP}$ is equal to $-1/3\chi_0$ at both band edges ($\mu = -6t, 3t$). 
Note that, at the top of the band, the effective mass is $m^*=2\hbar^2/3a^2 t$, which 
is twice larger than that at the bottom of the band. 
However, there are two hole pockets around the $K$ and $K'$ points, and thus 
$\chi_{\rm LP}(\mu=3t)$ is equal to $\chi_{\rm LP}(\mu=-6t)$. 

(2) $\chi_{\rm LP}$ increases as $\mu$ increases from $\mu=-6t$, 
crosses zero at $\mu=-0.190t$, and has a diverging peak at $\mu=2t$ corresponding 
to the van Hove singularities at ${\bm k}= (\pm \pi/a, \pm \sqrt{3}\pi/3a)$. 
This divergence is given by 
\begin{equation} 
\chi_{\rm LP}(\mu\rightarrow 2t) \sim \frac{\sqrt{3}}{3\pi} \ln \left( \frac{8t}{|\mu-2t|}\right) \ \chi_0,
\end{equation}
from the analytical form of $\chi_{\rm LP}$ in Appendix C. 
Here, we have used the fact that $\kappa$ in the elliptic functions such as (\ref{TriangleDOS}) 
behaves as $\kappa \sim 1+(\mu-2t)^3/64t^3$ near $\mu\le 2t$. 
This divergence is $\frac{1}{2}\frac{e^2}{\hbar^2 c^2} a^4 t^2 L^2$ times larger than the 
divergence of the density of states, $D(\mu)$ in (\ref{TriangleDOS}), which is 
reasonable since the integrand of $\chi_{\rm LP}$ is 
$\varepsilon_{xx} \varepsilon_{yy} - \varepsilon_{xy}^2 = 3t^2 a^4$ at the van Hove singularities. 

(3) In order to study the relative weights, we can choose the 
typical values for $\chi_{\rm FS}$ and $\chi_{{\rm occ}:2,3}$  
at the van Hove singularity ($\mu=2t$) similarly to the case of the square lattice. 
They are given by 
\begin{equation}
\chi_{\rm FS}(\mu=2t) = -\frac{16\sqrt{3}}{3\pi} \frac{1}{p^2} 
\left( 1+ b_1 \right) \chi_0, 
\end{equation}
\begin{equation}
\chi_{{\rm occ}:2,3} (\mu=2t) = \frac{8\sqrt{3}}{3\pi} \frac{1}{p^2} 
\left( 1-b_2+\frac{4}{5} b_1 \right) \chi_0,
\end{equation}
which are similar to the square lattice case. 
Therefore, the relative weights in the triangular-lattice case are also similar 
to those shown in Fig.~\ref{Fig:SquareRel}.

\section{Summary and Discussion}

We have calculated the orbital susceptibility in a single-band model  
up to the first-order with respect to overlap integrals between neighboring atomic orbitals. 
All the contributions including the deformation of Bloch wave functions due to magnetic 
field are included. 

In the zeroth order, we obtain the contribution of {\it intraband atomic diamagnetism}, 
$\chi_{{\rm occ}:1}$, which is proportional to the electron number in the partially filled band. 
This contribution gives asymmetry of the total susceptibility as a function of $\mu$ 
in the square-lattice case, which has not been recognized before. 
Since the other contributions are in the first order of overlap integrals, 
the relative weight of $\chi_{{\rm occ}:1}$ becomes larger 
as the atomic distance ($p=a/a_{\rm B}$) increases. 

In the first order with respect to overlap integrals, there are contributions 
from the Fermi surface ($\chi_{\rm FS}$) and from the occupied states 
($\chi_{{\rm occ}:2,3}$) in addition to the Landau-Peierls orbital susceptibility 
($\chi_{\rm LP}$). 
They also give comparable contributions as $\chi_{\rm LP}$, although their 
relative weights decrease as $p$ increases. 
It is known that the Peierls phase gives only $\chi_{\rm LP}$ in the single-band model.\cite{Peierls,Piechon}
Therefore, the present result means that the Peierls phase is insufficient 
as the effect of magnetic field. 
From the present derivation, it is apparent that the deformation of the 
wave function, $\Uell({\bm r})$, due to the magnetic field plays important roles 
that lead to additional contributions to $\chi_{\rm LP}$. 
The origin of this failure of the Peierls phase will be studied further in a separate paper.\cite{Matsuura} 

Here, we compare the present result with the previous results. 
As discussed by Raoux et al.,\cite{Piechon} when one restricts the band indices of the 
Green's functions in the Fukuyama formula (\ref{FukuyamaF}) to a single band, 
one obtains a susceptibility
\begin{equation}
\begin{split}
\chi_1 = \frac{e^2}{6\hbar^2 c^2} \sum_{\ell, {\bm k}} &f'(\varepsilon_\ell) \biggl\{  
\frac{\partial^2 \varepsilon_\ell}{\partial k_x^2} 
\frac{\partial^2 \varepsilon_\ell}{\partial k_y^2} + 2 \left(
\frac{\partial^2 \varepsilon_\ell}{\partial k_x \partial k_y} \right)^2  \cr
&+ \frac{3}{2} \left( 
\Eellx \frac{\partial^3 \varepsilon_\ell}{\partial k_x \partial k_y^2} + 
\Eelly \frac{\partial^3 \varepsilon_\ell}{\partial k_x^2 \partial k_y}
\right) \biggr\},
\label{Chi1}
\end{split}
\end{equation}
which is the same as $\chi_1$ used in I.\cite{OgaFuku1}
This $\chi_1$ is different from $\chi_{\rm LP}$ and also from the present result. 
This is natural since we have shown in I that there 
are other contributions in addition to $\chi_1$. 
This means that the band indices of the Green's functions in the Fukuyama formula 
should not be restricted to a single band.\cite{OgaFuku1}

Furthermore, Raoux et al.\cite{Piechon} compared $\chi_{\rm LP}$ and the first 
term of Hebborn et al.\cite{HS3}
\begin{equation}
\begin{split}
\chi_1^{({\rm HLSS})} = \frac{e^2}{6\hbar^2 c^2} &\sum_{\ell, {\bm k}} f'(\Eell) 
\biggl\{ \frac{\partial^2 \Eell}{\partial k_x^2} \frac{\partial^2 \Eell}{\partial k_y^2} 
-\left( \frac{\partial^2 \Eell}{\partial k_x \partial k_y} \right)^2 \cr
&+ \frac{3}{2} \left( \Eellx \frac{\partial^3 \Eell}{\partial k_x \partial k_y^2} 
+ \Eelly \frac{\partial^3 \Eell}{\partial k_x^2 \partial k_y} \right) \biggr\}. 
\end{split}
\end{equation}
This $\chi_1^{({\rm HLSS})}$ is also different from the present result. 
Since our formula is equivalent to that in Ref.~\cite{HS3}, our result should be obtained 
when we calculate all the contributions $\chi_1^{({\rm HLSS})}$-$\chi_4^{({\rm HLSS})}$ 
of Ref.~\cite{HS3}. 
This means that the other terms, i.e., $\chi_2^{({\rm HLSS})}, \chi_3^{({\rm HLSS})}$, 
and $\chi_4^{({\rm HLSS})}$ give comparable contributions and thus should not be neglected. 

In this paper, we calculate $\chi$ exactly up to the first order of overlap integrals. 
It is straightforward to study higher-order terms, but many contributions will appear. 
It will also be possible to calculate (\ref{FinalChi})-(\ref{ChiC}) numerically using the 
wave function (LCAO) of eq.~(\ref{LCAO}), where $\Phi_n({\bm r}-{\bm R}_i)$ includes 
higher order of overlap integrals. 
This is left as an interesting future problem. 

In the present 1s orbital case, $\chi_{\rm inter}$ vanishes because $L_z \phi_{\rm 1s}({\bm r})=0$
holds. 
It is interesting to study the cases in which $\chi_{\rm inter}$ has a finite contribution. 
For example, a model of graphene (or two-dimensional honeycomb lattice) 
is a typical two-band model. 
In this case, $\chi_{\rm inter}$ can have a finite contribution  
even in the first order with respect to overlap integrals. 
Calculations based on the exact formula will be published in a following paper. 

\bigskip\noindent
{\bf Acknowledgment}

We thank H.\ Fukuyama, F.\ Pi\'echon, H.\ Matsuura, 
I.\ Proskurin, T.\ Kariyado, Y.\ Fuseya, T.\ Mizoguchi, and N.\ Okuma for very fruitful discussions.  
This work was supported by a Grant-in-Aid for Scientific Research on 
\lq\lq Multiferroics in Dirac electron materials'' (No.\ 15H02108).

\appendix

\section{Overlap integrals}

The overlap integrals are defined in eq.~(\ref{Odefinition}). 
First, note that there is a difference between $\Phi_{\ell}({\bm r})$ 
and $\phi_{\ell}({\bm r})$ as in eq.~(\ref{OrthoNormal}). 
Therefore, we introduce expectation values in terms of 
$\phi_{\ell}({\bm r})$ as follows:
\begin{equation}
\begin{split}
\langle {\cal O} \rangle_{\ell\ell}^{(0)} 
&= \int \phi_\ell^* ({\bm r}) {\cal O} \phi_{\ell} ({\bm r}) d{\bm r}, \cr
\langle {\cal O} \rangle_{R\ell\ell}^{(0)} 
&= \int \phi_\ell^* ({\bm r}-{\bm R}) {\cal O} \phi_{\ell} ({\bm r}) d{\bm r}.
\end{split}
\end{equation}
The expectation values in terms of $\Phi_{\ell}({\bm r})$ can be easily obtained from these values. 

First, we prove some exact equalities that hold quite generally. 
In the case where the atomic orbital $\phi_\ell ({\bm r})$ satisfies
$\phi_\ell (-{\bm r})=\pm \phi^*_\ell ({\bm r})$, we obtain
\begin{equation}
\begin{split}
\langle x \rangle_{R\ell\ell}^{(0)} 
&\equiv \int \phi_\ell^* ({\bm r}-{\bm R}) x \phi_{\ell} ({\bm r}) d{\bm r} \cr
&= \int \phi_\ell^* (-{\bm r}') (-x'+R_x) \phi_{\ell} (-{\bm r}'+{\bm R}) d{\bm r}' \cr
&= \int \phi_\ell ({\bm r}) (-x+R_x) \phi^*_{\ell} ({\bm r}-{\bm R}) d{\bm r} \cr
&= -\langle x \rangle_{R\ell\ell}^{(0)} + R_x \langle 1 \rangle_{R\ell\ell}^{(0)} = \frac{R_x}{2} s, 
\label{Formula1}
\end{split}
\end{equation}
\begin{equation}
\begin{split}
\langle p_x x \rangle_{R\ell\ell}^{(0)} 
&\equiv \int \phi_\ell^* ({\bm r}-{\bm R}) p_x x \phi_{\ell} ({\bm r}) d{\bm r} \cr
&= \int \phi_\ell^* (-{\bm r}') (-p_{x'}) (-x'+R_x) \phi_{\ell} (-{\bm r}'+{\bm R}) d{\bm r}' \cr
&= -\langle xp_x \rangle_{R\ell\ell}^{(0)} + R_x \langle p_x \rangle_{R\ell\ell}^{(0)}  \cr
&= -\langle p_x x \rangle_{R\ell\ell}^{(0)} - i\hbar \langle 1 \rangle_{R\ell\ell}^{(0)} 
+ R_x \langle p_x \rangle_{R\ell\ell}^{(0)} \cr
&= -\frac{i\hbar}{2} s + \frac{R_x}{2} \langle p_x \rangle_{R\ell\ell}^{(0)}, 
\end{split}
\end{equation}
where we have used the change of the variable ${\bm r}=-{\bm r}'+{\bm R}$ 
and $s=\langle 1 \rangle_{R\ell\ell}^{(0)}$. 
In a similar way, we can prove
\begin{equation}
\begin{split}
\langle y \rangle_{R\ell\ell}^{(0)} &= \frac{R_y}{2} s, \quad
\langle p_y x \rangle_{R\ell\ell}^{(0)} = \frac{R_x}{2} \langle p_y \rangle_{R\ell\ell}^{(0)}, \cr
\langle p_x y \rangle_{R\ell\ell}^{(0)} &= \frac{R_y}{2} \langle p_x \rangle_{R\ell\ell}^{(0)}, \quad
\langle p_y y \rangle_{R\ell\ell}^{(0)} = -\frac{i\hbar}{2} s + \frac{R_y}{2} \langle p_y \rangle_{R\ell\ell}^{(0)}. 
\label{Formula2}
\end{split}
\end{equation}
When ${\bm R}=0$, we can also show 
$\langle x \rangle_{\ell\ell}^{(0)}=\langle y \rangle_{\ell\ell}^{(0)}=\langle p_y x \rangle_{\ell\ell}^{(0)}=0$ 
and $\langle p_x x \rangle_{\ell\ell}^{(0)}=\langle p_y y \rangle_{\ell\ell}^{(0)}=-i\hbar/2$, etc.

Next, when the atomic orbital $\phi_\ell ({\bm r})$ is isotropic in three-dimensional 
space like the 1s orbital, or 
when it is isotropic in the $xy$-plane like the ${\rm p}_\pi$ orbital, we can prove
\begin{equation}
\begin{split}
\langle p_x \rangle_{R\ell\ell}^{(0)} &= \frac{R_x}{a} \langle p_\parallel \rangle_{R\ell\ell}^{(0)}, \quad
\langle p_y \rangle_{R\ell\ell}^{(0)} = \frac{R_y}{a} \langle p_\parallel \rangle_{R\ell\ell}^{(0)}, \cr
\langle x^2   \rangle_{R\ell\ell}^{(0)} &= \left( 1-\frac{R_x^2}{a^2} \right) \langle r_\perp^2 \rangle_{R\ell\ell}^{(0)}
+ \frac{R_x^2}{a^2} \langle r_\parallel^2 \rangle_{R\ell\ell}^{(0)}, \cr
\langle y^2   \rangle_{R\ell\ell}^{(0)} &= \left( 1-\frac{R_y^2}{a^2} \right) \langle r_\perp^2 \rangle_{R\ell\ell}^{(0)}
+ \frac{R_y^2}{a^2} \langle r_\parallel^2 \rangle_{R\ell\ell}^{(0)}, 
\label{Formula3}
\end{split}
\end{equation}
with the help of the rotation of the coordinates. 
Here,  $a=|{\bm R}|$ and $p_\parallel$ represents the momentum operator 
in the direction parallel to $\bm R$, 
while $r_\perp (r_\parallel)$ means the coordinate in the direction perpendicular (parallel) 
to $\bm R$. 
Note that $\langle p_\perp \rangle_{R\ell\ell}^{(0)}=0$ from symmetry. 
Furthermore, we can show that $L_z \phi_{\ell}({\bm r})=0$ and 
\begin{equation}
\langle L_z y \rangle_{R\ell\ell}^{(0)} = 
\langle y L_z -i\hbar x \rangle_{R\ell\ell}^{(0)} = -\frac{i\hbar}{2} sR_x,
\label{FormulaLzy}
\end{equation}
where a commutation relation, $[L_z , y] = -i\hbar x$, has been used. 

Various kinds of integrals can be carried out explicitly when we use the atomic orbitals. 
Without loss of generality, we assume ${\bm R}=(a,0,0)$. 
Then, by using a change of coordinates, 
$\xi= r+r_b, \eta = r-r_b$ with $r=|{\bm r}|, r_b = |{\bm r}-{\bm R}|$,\cite{Mulliken} 
we obtain for the 1s orbital 
\begin{equation}
\begin{split}
\langle 1 \rangle_{R{\rm 1s1s}}^{(0)} &= \left( 1+p+\frac{p^2}{3} \right)\ e^{-p}, \cr
\langle \frac{1}{r} \rangle_{R{\rm 1s1s}}^{(0)} &= \frac{1}{a_{\rm B}} \left( 1+p \right)\ e^{-p}, \cr
\langle p_\parallel \rangle_{R{\rm 1s1s}}^{(0)} &= \frac{i\hbar}{3 a_{\rm B}} p \left( 1+p \right) \ e^{-p}, \cr
\langle r_\perp^2 \rangle_{R{\rm 1s1s}}^{(0)} &= 
a_{\rm B}^2 \left( 1+p+\frac{2}{5} p^2 + \frac{p^3}{15} \right)\ e^{-p}, \cr 
\langle r_\parallel^2 \rangle_{R{\rm 1s1s}}^{(0)} &= 
a_{\rm B}^2 \left( 1+p+\frac{7}{10} p^2 + \frac{11}{30} p^3 + \frac{p^4}{10} \right)\ e^{-p},
\label{IntResult1s}
\end{split}
\end{equation}
with $p=a/a_{\rm B}$. 
The first two equations give $s$ and $t_0$ in eq.~(\ref{OverlapInt1s}). 
When we put $p=0$ in the last two equations, we obtain 
$\langle x^2 \rangle_{\rm 1s1s}^{(0)}=\langle y^2 \rangle_{\rm 1s1s}^{(0)}=a_{\rm B}^2$. 

Finally, we calculate the expectation values in terms of $\Phi_{\ell} ({\bm r})$. 
Using the relation (\ref{OrthoNormal1s}) for the 1s case, we can show 
\begin{equation}
\langle {\cal O} \rangle_{\ell\ell} = \langle {\cal O} \rangle_{\ell\ell}^{(0)} + O(s^2).
\end{equation}
Therefore, up to the first order of overlap integrals, $\langle {\cal O} \rangle_{\ell\ell}$
and $\langle {\cal O} \rangle_{\ell\ell}^{(0)}$ are equivalent. 
For $\langle {\cal O} \rangle_{R\ell\ell}$, we can show
\begin{equation}
\begin{split}
\langle {\cal O}({\bm r}) \rangle_{R\ell\ell} 
&= \langle {\cal O}({\bm r}) \rangle_{R\ell\ell}^{(0)} -\frac{s}{2} \sum_{{\bm R}'} 
\int \phi_\ell^* ({\bm r}-{\bm R}+{\bm R}') {\cal O}({\bm r}) \phi_{\ell} ({\bm r}) d{\bm r} \cr
&\qquad-\frac{s}{2} \sum_{{\bm R}'} 
\int \phi_\ell^* ({\bm r}-{\bm R}) {\cal O}({\bm r}) \phi_{\ell} ({\bm r}-{\bm R}') d{\bm r} +O(s^2) \cr
&= \langle {\cal O}({\bm r}) \rangle_{R\ell\ell}^{(0)} -\frac{s}{2} \langle {\cal O}({\bm r}) \rangle_{\ell\ell}^{(0)}
-\frac{s}{2} \langle {\cal O}({\bm r}+{\bm R}) \rangle_{\ell\ell}^{(0)} + O(s^2), 
\end{split}
\end{equation}
with ${\bm R}' = {\bm R}_j - {\bm R}_i$.  
Here, we have taken into account only the term with ${\bm R}'={\bm R}$
in the ${\bm R}'$-summation in the first-order of the overlap integrals. 

Using this relation and (\ref{Formula1})-(\ref{FormulaLzy}), we can show 
 \begin{equation}
\begin{split}
\langle 1 \rangle_{R\ell\ell} &=
\langle x \rangle_{R\ell\ell} = \langle y \rangle_{R\ell\ell}=0, \cr
\langle p_y y \rangle_{R\ell\ell} &= \frac{R_y}{2} \langle p_y \rangle_{R\ell\ell}^{(0)} 
=\frac{R_y^2}{2a} \langle p_\parallel \rangle_{R\ell\ell}^{(0)}, \cr
\langle p_x y \rangle_{R\ell\ell} &= \frac{R_y}{2} \langle p_x \rangle_{R\ell\ell}^{(0)} 
=\frac{R_x R_y}{2a} \langle p_\parallel \rangle_{R\ell\ell}^{(0)}, \cr
\langle L_z y \rangle_{R\ell\ell} &= -\frac{i\hbar}{2} sR_x 
-\frac{sR_x}{2} \langle p_y y \rangle_{\ell\ell}^{(0)} =-\frac{i\hbar}{4} sR_x,  
\label{FormulaRll}
\end{split}
\end{equation}
and 
\begin{equation}
\langle x^2 \rangle_{R\ell\ell} = \langle x^2 \rangle_{R\ell\ell}^{(0)} 
- s \langle x^2 \rangle_{\ell\ell}^{(0)} -\frac{s}{2} R_x^2. 
\end{equation}
Substituting the explicit integrals in (\ref{IntResult1s}) 
for the 1s orbital, we obtain 
\begin{equation}
\langle x^2 \rangle_{R{\rm 1s1s}} = \frac{a^2}{15} (1+p)\ e^{-p} 
-\frac{R_x^2}{5} \left( 1+p +\frac{p^2}{3}\right) \ e^{-p}.
\label{x2in1s}
\end{equation}
There is a relation
\begin{equation}
\langle x^2+y^2 \rangle_{R{\rm 1s1s}}^{(0)} 
- s \langle x^2+y^2\rangle_{{\rm 1s1s}}^{(0)} = \frac{mta^2a_{\rm B}^2}{\hbar^2}
\left( \frac{6}{5} b_1+b_2 \right), 
\end{equation}
which can be used in $\chi_{\rm FS}+\chi_{\rm occ}$. 

\section{${\bm k}$-integrals for square lattice}

In the case of square lattice, the density of states per area is given by
\begin{equation}
\begin{split}
D(\mu) &\equiv \frac{1}{L^2} \sum_{\bm k} \delta(\varepsilon_{\bm k}-\mu) \cr
&= \frac{1}{(2\pi)^2 a^2} \iint dk_x dk_y 
\delta( -2t(\cos k_x +\cos k_y) -\mu),
\end{split}
\end{equation}
with $L^2=Na^2$. 
We find that it is convenient to use the variables $v=\cos k_x + \cos k_y$ and 
$u=\cos k_x-\cos k_y$.
Then, it is straightforward to obtain the density of states per area as 
\begin{equation}
\begin{split}
D(\mu) &= \frac{1}{2\pi^2 a^2} \int_{-2}^{2} du \int_{-2+|u|}^{2-|u|} dv
\frac{\delta( -2tv -\mu)}{\sqrt{1-\left( \frac{u+v}{2} \right)}^2 \sqrt{1-\left( \frac{u-v}{2} \right)}^2} \cr
&=\frac{1}{\pi^2 (1+k') ta^2} K\left( \frac{1-k'}{1+k'} \right) \cr
&=\frac{1}{2\pi^2 ta^2} K(k),
\end{split}
\end{equation}
where $K(k)$ is the complete elliptic integral of the first kind, 
$k'=|\mu|/4t$, and $k=\sqrt{1-k'^2}= \sqrt{1-\mu^2/16t^2}$. 

For $\chi_{\rm LP}$, we need to calculate an integral with $\cos k_x \cos k_y$. 
By the same method as in the density of states, we obtain
\begin{equation}
\begin{split}
&\frac{1}{L^2} \sum_{\bm k} \delta(\varepsilon_{\bm k}-\mu) (\varepsilon_{xx} \varepsilon_{yy} - \varepsilon_{xy}^2) \cr
&=\frac{4t^2a^2}{(2\pi)^2} \iint dk_x dk_y \delta( -2t(\cos k_x +\cos k_y) -\mu) \cos k_x \cos k_y \cr
&= \frac{t^2a^2}{2\pi^2} \int_{-2}^{2} du \int_{-2+|u|}^{2-|u|} dv
\frac{\delta( -2tv -\mu)(v^2-u^2)}{\sqrt{1-\left( \frac{u+v}{2} \right)}^2 \sqrt{1-\left( \frac{u-v}{2} \right)}^2} \cr
&=\frac{4ta^2}{\pi^2} \left\{ (1+k') E\left( \frac{1-k'}{1+k'} \right) 
-\frac{1+2k'}{1+k'} K\left( \frac{1-k'}{1+k'} \right) \right\} \cr
&=\frac{4ta^2}{\pi^2} \left\{ E(k)-\frac{1}{2} K(k) \right\}.
\end{split}
\end{equation}
Similarly, for $\chi_{\rm FS}$, we obtain
\begin{equation}
\begin{split}
&\frac{1}{L^2} \sum_{\bm k} \delta(\varepsilon_{\bm k}-\mu) (\varepsilon_{x}^2 + \varepsilon_{y}^2) \cr
&=\frac{4t^2}{(2\pi)^2} \iint dk_x dk_y \delta( -2t(\cos k_x +\cos k_y) -\mu) (\sin^2 k_x + \sin^2 k_y) \cr
&= \frac{2t^2}{\pi^2} \int_{-2}^{2} du \int_{-2+|u|}^{2-|u|} dv
\frac{\delta( -2tv -\mu)(2-\frac{v^2}{2}-\frac{u^2}{2})}
{\sqrt{1-\left( \frac{u+v}{2} \right)}^2 \sqrt{1-\left( \frac{u-v}{2} \right)}^2} \cr
&=\frac{8t}{\pi^2} \left\{ (1+k') E\left( \frac{1-k'}{1+k'} \right) 
-2k' K\left( \frac{1-k'}{1+k'} \right) \right\} \cr
&=\frac{8t}{\pi^2} \left\{ E(k)- (1-k^2) K(k) \right\}.
\end{split}
\end{equation}

\def\Myeta{\eta}

\section{${\bm k}$-integrals for triangular lattice}

In the case of a triangular lattice, there is no useful trick for the $\bm k$-integrals 
as in the square lattice. 
In this case, the Brillouin zone is a honeycomb with a size of $8\sqrt{3}\pi^2/3a^2$ 
with $a$ being the nearest-neighbor distance, and the system area is $L^2=\sqrt{3}a^2 N/2$ 
with $N$ being the total number of sites. 
The density of states per area is obtained as
\begin{equation}
\begin{split}
D(\mu) &= \frac{2\sqrt{3}}{3Na^2} \sum_{\bm k} \delta(\varepsilon_{\bm k}-\mu) \cr 
&= \frac{1}{(2\pi)^2a^2} \iint_{\rm B.Z.} dk_x dk_y \cr
&\quad \times \delta \left( -2t ( \cos k_x + 2\cos \frac{k_x}{2} \cos \frac{\sqrt{3}k_y}{2} ) -\mu\right).
\label{TriangleIntDOS}
\end{split}
\end{equation}
After the $k_y$-integral and a change of the variable $x=\cos^2 k_x/2$, we obtain
\begin{equation}
\begin{split}
D(\mu) &=\frac{1}{\sqrt{3}\pi^2 ta^2} \int_{0}^{1} 
\frac{\theta( 4x - (2x-1+\mu/2t)^2)dx}{\sqrt{x}\sqrt{1-x}\sqrt{4x - (2x-1+\mu/2t)^2}}\cr
&=\frac{1}{2\sqrt{3}\pi^2 ta^2} \int_{0}^{1} 
\frac{\theta( (\alpha-x)(x-\beta) )}{\sqrt{x(1-x)(\alpha-x)(x-\beta)}} dx,
\end{split}
\end{equation}
where $\theta(x)$ is a step function [$\theta(x)=1, x>0$ and $\theta(x)=0, x<0$], 
$\alpha = (1+\Myeta)^2/4, \beta = (1-\Myeta)^2/4$, 
and $\Myeta=\sqrt{3-\mu/t}$.
Finally, using the formula\cite{GRx}
\begin{equation}
\int_{c}^{b} \frac{dx}{\sqrt{(a-x)(b-x)(x-c)(x-d)}} = \frac{2}{\sqrt{(a-c)(b-d)}} K(q),
\label{Formula10}
\end{equation}
for $a>b>c>d$ with  
\begin{equation}
q = \sqrt{\frac{(a-d)(b-c)}{(a-c)(b-d)}},
\end{equation}
we obtain eq.~(\ref{TriangleDOS}). 

For $\chi_{\rm LP}$ of the triangular lattice, we need to calculate the integral
\begin{equation}
\begin{split}
I_1(\mu) &= \frac{1}{L^2} \sum_{\bm k} \delta(\varepsilon_{\bm k}-\mu) 
(\varepsilon_{xx} \varepsilon_{yy} - \varepsilon_{xy}^2) \cr
&= \frac{3t^2a^2}{(2\pi)^2} \iint_{\rm B.Z.} dk_x dk_y \delta(\varepsilon_{\bm k}-\mu) \cr
&\times \biggr\{ 2\cos k_x \cos \frac{k_x}{2} \cos \frac{\sqrt{3}k_y}{2} \cr
&+ \cos^2 \frac{k_x}{2} \cos^2 \frac{\sqrt{3}k_y}{2} -  \sin^2 \frac{k_x}{2} \sin^2 \frac{\sqrt{3}k_y}{2} \biggr\}.
\end{split}
\end{equation}
After some algebra, this integral can be rewritten as
\begin{equation}
\begin{split}
I_1(\mu) &=\frac{\sqrt{3} ta^2}{2\pi^2} \int_{0}^{1} 
\frac{\theta( (\alpha-x)(x-\beta) )}{\sqrt{x(1-x)(\alpha-x)(x-\beta)}}\cr
&\times \left\{ -4x^2 + \left( 6-\frac{\mu}{t} \right) x -3 + \frac{\mu}{t} 
+ \frac{1}{4x} \left( 1-\frac{\mu}{2t} \right)^2 \right\} dx.
\label{TriangleIntLP}
\end{split}
\end{equation}
In order to perform the $x$-integral, we can use the formula
\begin{equation}
\begin{split}
&\int_{c}^{b} \frac{(x-d) dx}{\sqrt{(a-x)(b-x)(x-c)(x-d)}} \cr
&= \frac{2(c-d)}{\sqrt{(a-c)(b-d)}} \Pi (\frac{\pi}{2}, \frac{b-c}{b-d},q),
\label{Formula11}
\end{split}
\end{equation}
\begin{equation}
\begin{split}
&\int_{c}^{b} \frac{(x-d)^2 dx}{\sqrt{(a-x)(b-x)(x-c)(x-d)}} \cr
&= \frac{(c-d)(a+b+c-3d)}{\sqrt{(a-c)(b-d)}} \Pi (\frac{\pi}{2}, \frac{b-c}{b-d},q) \cr
&- \sqrt{(a-c)(b-d)} E(q) - \frac{(b-d)(c-d)}{\sqrt{(a-c)(b-d)}} K(q), 
\label{Formula12}
\end{split}
\end{equation}
\begin{equation}
\begin{split}
&\int_{c}^{b} \frac{dx}{(x-d)\sqrt{(a-x)(b-x)(x-c)(x-d)}} \cr
&= \frac{2}{(b-d) \sqrt{(a-c)(b-d)}} \left\{ \frac{(a-b)q}{a-d} \frac{dK}{dk}\biggr|_{k=q} +K(q) \right\} ,
\label{Formula13}
\end{split}
\end{equation}
with 
\begin{equation}
\begin{split}
\Pi (\varphi, n, q)=\int_0^{\varphi} \frac{d\theta}{(1-n\sin^2 \theta)\sqrt{1-k^2 \sin^2 \theta}},
\end{split}
\end{equation}
being the elliptic integral of the third kind. 
(\ref{Formula11}) and (\ref{Formula12}) are obtained by integrating the parameter $a$ in the formula 
(\ref{Formula10}) from $b$ to $a$, i.e., $\int_b^a \cdots da$, while (\ref{Formula13}) 
is obtained by differentiating with respect to the parameter $d$. 

Using these formulas, (\ref{TriangleIntLP}) becomes
\begin{equation}
\begin{split}
I_1(\mu) &=\frac{\sqrt{3} ta^2}{2\pi^2} \biggl[ 4\sqrt{\eta} E(\kappa)
+ \frac{\frac{\mu^2}{t^2}+\frac{4\mu}{t}-12-16\eta}{8\sqrt{\eta}} K(\kappa) \cr
&+ \frac{\frac{\mu^2}{t^2}-12+8\eta}{8\sqrt{\eta}} \kappa \frac{dK}{dk}\biggr|_{k=\kappa} \biggr],
\label{TriangleIntLP2}
\end{split}
\end{equation}
for $-6<\mu/t<2$ and
\begin{equation}
\begin{split}
I_1(\mu) &=\frac{\sqrt{3} ta^2}{2\pi^2} \biggl[ 4\kappa \sqrt{\Myeta} E(\frac{1}{\kappa})
+ \frac{\frac{\mu^2}{t^2}+\frac{2\mu}{t}-12-4\eta}{4\kappa \sqrt{\Myeta}} K(\frac{1}{\kappa}) \cr
&- \frac{\frac{\mu^2}{t^2}-12+8\eta}{8\kappa \sqrt{\Myeta}} \frac{1}{\kappa} \frac{dK}{dk}\biggr|_{k=1/\kappa} \biggr],
\label{TriangleIntLP3}
\end{split}
\end{equation}
for $2<\mu/t<3$, where 
\begin{equation}
\kappa = \sqrt{\frac{\alpha(1-\beta)}{\Myeta}} 
= \sqrt{\frac{-\frac{\mu^2}{t^2}+12+8\Myeta}{\Myeta}}.
\end{equation}

For $\chi_{\rm FS}$, we need to calculate an integral with 
\begin{equation}
\left( \sin k_x + \sin \frac{k_x}{2} \cos \frac{\sqrt{3}k_y}{2} \right)^2 
+ 3\cos^2 \frac{k_x}{2} \sin^2 \frac{\sqrt{3}k_y}{2},
\end{equation}
which comes from $\varepsilon_x^2+\varepsilon_y^2$. 
Then the integral becomes
\begin{equation}
\begin{split}
I_2(\mu) &= \frac{1}{L^2} \sum_{\bm k} \delta(\varepsilon_{\bm k}-\mu) (\varepsilon_{x}^2+\varepsilon_{y}^2) \cr
&=\frac{2t}{\sqrt{3}\pi^2} \int_{0}^{1} 
\frac{\theta( (\alpha-x)(x-\beta) )}{\sqrt{x(1-x)(\alpha-x)(x-\beta)}}\cr
&\times \left\{ -4x^2 + \left( 6-\frac{\mu}{t} \right) x + \frac{\mu}{2t}-\frac{\mu^2}{4t^2}  
+ \frac{1}{4x} \left( 1-\frac{\mu}{2t} \right)^2 \right\} dx.
\label{TriangleIntFS}
\end{split}
\end{equation}
Therefore, we obtain
\begin{equation}
\begin{split}
I_2(\mu) &=\frac{2t}{\sqrt{3}\pi^2} \biggl[ 4\sqrt{\eta} E(\kappa)
- \frac{\frac{3\mu^2}{t^2}+\frac{4\mu}{t}-36+16\eta}{8\sqrt{\eta}} K(\kappa) \cr
&+ \frac{\frac{\mu^2}{t^2}-12+8\eta}{8\sqrt{\eta}} \kappa \frac{dK}{dk}\biggr|_{k=\kappa} \biggr],
\label{TriangleIntFS2}
\end{split}
\end{equation}
for $-6<\mu/t<2$ and
\begin{equation}
\begin{split}
I_2(\mu) &=\frac{2t}{\sqrt{3}\pi^2} \biggl[ 4\kappa \sqrt{\Myeta} E(\frac{1}{\kappa})
- \frac{\frac{\mu^2}{t^2}+\frac{2\mu}{t}-12+4\eta}{4\kappa \sqrt{\Myeta}} K(\frac{1}{\kappa}) \cr
&- \frac{\frac{\mu^2}{t^2}-12+8\eta}{8\kappa \sqrt{\Myeta}} \frac{1}{\kappa} \frac{dK}{dk}\biggr|_{k=1/\kappa} \biggr],
\label{TriangleIntFS3}
\end{split}
\end{equation}
for $2<\mu/t<3$.

\def\journal#1#2#3#4{#1 {\bf #2}, #3 (#4)}
\def\PR{Phys.\ Rev.}
\def\PRB{Phys.\ Rev.\ B}
\def\PRL{Phys.\ Rev.\ Lett.}
\def\JPSJ{J.\ Phys.\ Soc.\ Jpn.}
\def\PTP{Prog.\ Theor.\ Phys.}
\def\JPCS{J.\ Phys.\ Chem.\ Solids}

\end{document}